\documentclass[twocolumn,pra,superscriptaddress]{revtex4}
\usepackage[colorlinks,linkcolor=blue,anchorcolor=blue,citecolor=blue,filecolor=blue,menucolor=blue,runcolor=blue,urlcolor=blue,frenchlinks=blue]{hyperref}
\usepackage{amsmath}
\usepackage{graphicx}
\usepackage{amssymb}
\usepackage{dcolumn}
\usepackage{mathrsfs}
\usepackage{bm}
\usepackage{color}
\usepackage{soul}
\usepackage{xcolor}
\begin{document}

\title{Can the gate time of Rydberg two-qubit gate be shorten by increasing the strength of Rydberg interaction?}

\author{Yan-Xiong Du}
\email{yanxiongdu@m.scnu.edu.cn}
\affiliation {Key Laboratory of Atomic and Subatomic Structure and Quantum Control (Ministry of Education), Guangdong Basic Research Center of Excellence for Structure and Fundamental Interactions of Matter, School of Physics, South China Normal University, Guangzhou 510006, China}

\affiliation {Guangdong Provincial Key Laboratory of Quantum Engineering and Quantum Materials, Guangdong-Hong Kong Joint Laboratory of Quantum Matter, Frontier Research Institute for Physics, South China Normal University, Guangzhou 510006, China}


\begin{abstract}
In this manuscript we discuss the relationship between the gate time of Rydberg two-qubit gate and the Rydberg interaction strength. Different from the two-qbuit gates that directly depend on the interactions between the spins (or pseudo spins), the ones in atomic arrays are realized by driving the atoms to the Rydberg states. As a consequence, competition happens between shortening the gate time and decreasing the excitation probability of Rydberg states. For the case of Rydberg blockade, it is found that the gate time is irrelevant of the Rydberg interaction strength. In contrast, for the case of weak Rydberg interactions, the interactions will help to accelerate the two-qubit gates. This implies that the scheme of weak Rydberg interactions will be faster than the Rydberg blockade one to realize the two-qbuit gates under the same Rabi frequencies of Rydberg excitation. Through using the geometric control under the region of weak Rydberg interaction, fast and robust two-qubit gates can be achieved in atomic arrays, of which detailed discussion can be referred to the manuscript \textsl{arXiv:2412. 19193 (2024)}.
\end{abstract}

 \maketitle

\section{Introduction}
Realizing fast and robust quantum control is the key task for quantum computation, which will significantly decrease the influence from the decoherence and dephasing effects. In the last ten years, the atomic arrays with Rydberg interaction has become important platforms for the investigation in quantum computation \cite{Saffman2016,Henriet2020,Browaeys2020,Scholl2021,Ebadi2021,Kim2024}. High-fidelity quantum gates in such system have also been realized which exceed the quantum error correction threshold \cite{Levine2019,Madjarov2020,Fu2022,Evered2023,Bluvstein2024}. In most reported experiments \cite{Isenhower2010,Jau2016,Theis2016,Wilk2010,Zeng2017,Muller2014}, the Rydberg blockade effect is adopted \cite{Liu2020,Shi2018,Sun2020,Li2022,Pagano2022,Goerz2014,Maller2015}. The gate time of two-qubit gates distributes in the region of hundreds of nanosecond to microsecond, which is still close to the dephasing time ($T_2$) of the Rydberg states. One may wonder if there exist new protocol to shorten the two-qubit gate time, without needing to increase the power of Rydberg lasers. A straight forward idea will be increasing the interaction strength between the atoms (using higher principal numbers of Rydberg states and shorter spacing). However, it is found that the gate time of two-qubit gates with Rydberg blockade are irrelevant to the interaction strength as will be shown in the following. Although a smaller interaction strength may be adopted \cite{Jo2020}, is must overcome the problem of the thermal emotion of atoms since the interaction strength will be sensitive to the atomic spacing $R$ ($1/R^6$ scaling law).

In this manuscript we discuss the relationship between the gate time of Rydberg two-qubit gates and the Rydberg interaction strength. For the case of Rydberg blockade, it is found that the gate time is irrelevant of the Rydberg interaction strength. Meanwhile, for the case of weak Rydberg interaction, the Rydberg interaction will help to accelerate the two-qubit gate. Combining the geometric control and the composite pulses, fast and robust two-qubit gate can be achieved in atomic arrays with weak Rydberg interaction \cite{Ming2024}.

\begin{figure}[ptb]
\begin{center}
\includegraphics[width=8.5cm]{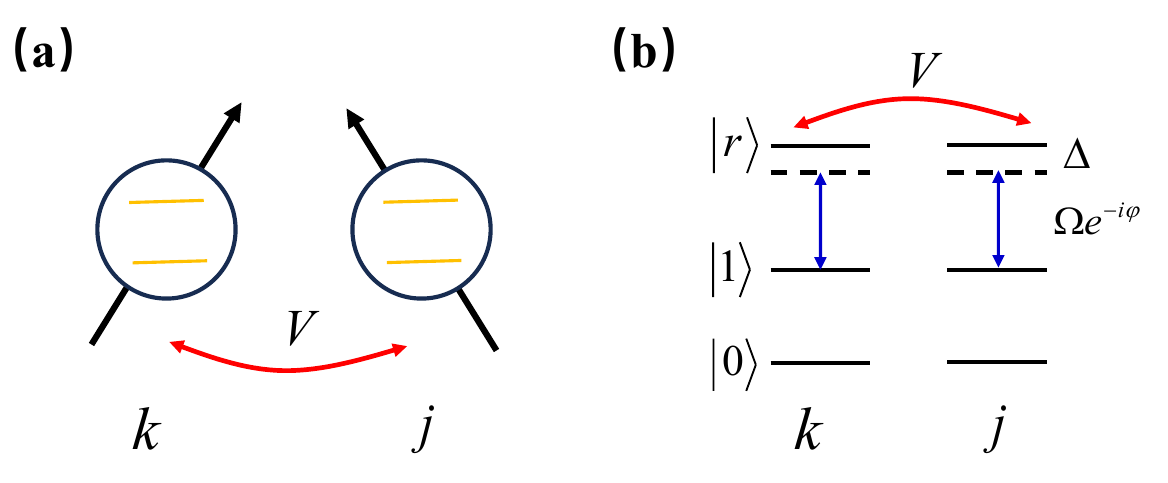}
\caption{Scheme of two-qubit gates. (a) Two-qubit gates with direct couplings. (b) Two-qubit gates with indirect couplings, i.e. through exciting to the Rydberg states.
}
\end{center}
\end{figure}

\section{Two-qubit gates with direct couplings}
We first briefly review the two-qubit gate with direct couplings as shown in Fig. 1(a). The interaction Hamiltonian between the two qubits $k, j$ is given by
\begin{equation}
H_{\mathrm{di}}=\sum_{a, b=x, y, z}J_{ab}\sigma_a^{k}\sigma_b^{j}.
\end{equation}
$J_{ab}$ is the interaction strength and $\sigma_{a,b}^{k,j}$ are the Pauli matrixes that defined in the computational basis of $k(j)$-th qubit. $H_{\mathrm{di}}$ can be realized by the spin interaction between the spins in NMR system, tunneling between two quantum dots or inductive coupling between two super-conducting qubits \cite{Du2000,Jones2012,You2002}. When XY-type interaction $H_{XY}=\sum_{k\neq j}J_{XY}(\sigma_x^{k}\sigma_x^{j}+\sigma_y^{k}\sigma_y^{j})$ is adopted, the evolution operator will be given by
\begin{equation}
U_{XY}=e^{-iH_{XY}t}=\left(\begin{array}{cccc}
         1&0&0&0  \\
         0&\cos\theta_{XY} & i\sin\theta_{XY} &0  \\
         0&i\sin\theta_{XY} & \cos\theta_{XY} &0  \\
         0&0&0&1
    \end{array}\right),
\end{equation}
where $\theta_{XY}=2J_{XY}t$. Controlled-flip gate based on $U_{XY}$ can be realized when $J_{XY}t=\pi/4$. When the ZZ-type interaction $H_{ZZ}=J_{ZZ}\sigma_Z^{k}\sigma_Z^{j}/4$ is adopted, we will derive the evolution operator as
\begin{equation}
U_{ZZ}=e^{-iH_{XY}t}=\left(\begin{array}{cccc}
         1&0&0&0  \\
         0&1 & 0 &0  \\
         0&0 & 1 &0  \\
         0&0&0&e^{-J_{ZZ}t}
    \end{array}\right).
\end{equation}
Controlled-phase gate based on $U_{ZZ}$ can be realized when $J_{ZZ}t=\pi$. Similar discussion can be found in the case with the interacting Hamiltonian of $H_{+-}=\sum_{k\neq j}J_{+-}(\sigma_+^{k}\sigma_-^{j}+\sigma_-^{k}\sigma_+^{j})$.

Therefore, it can be found that for the case of two-qubit gates with direct couplings, the gate time can be shorten by increasing the interacting strength.

\section{Two-qubit gates with strong Rydberg interactions}
In the following we investigate the two-qubit gates in atomic arrays, as shown in Fig. 1(b). Two atoms (labeled $k, j$) with computational basis $\{|0\rangle, |1\rangle\}$ are adopted. The interactions between two atoms are realized by driving the atoms to the Rydberg states $|r\rangle$, different from the former scheme of direct couplings. We assume that state $|1\rangle$ is coupled to $|r\rangle$ with Rabi frequency $\Omega$, detuning $\Delta$ and phase $\varphi$. The interaction strength between two Rydberg states is given by $V$. Under the single-qubit basis, the interacting Hamiltonian is given by
\begin{eqnarray}
&&H=(\frac{\Omega}{2}e^{i\varphi}|1\rangle_1\langle r|\otimes I_2+I_1\otimes|1\rangle_2\langle r|+\mathrm{H.c.})\\
&&+\Delta(|r\rangle_1\langle r|\otimes I_2+I_1\otimes|r\rangle_2\langle r|)+V|r\rangle_1\langle r|\otimes|r\rangle_2\langle r|, \nonumber
\end{eqnarray}
where $I_1=|0\rangle_1\langle 0|+|1\rangle_1\langle 1|+|r\rangle_1\langle r|$, $I_2=|0\rangle_2\langle 0|+|1\rangle_2\langle 1|+|r\rangle_2\langle r|$, $\hbar=1$. It can be checked that states $\{|01\rangle, |0r\rangle\}$, $\{|10\rangle, |r0\rangle\}$ and $\{|11\rangle, |R\rangle, |rr\rangle\}$ form subspace respectively and the corresponding Hamiltonian are given by
\begin{subequations}
\begin{align}
H_{01}&=(\frac{\Omega}{2}e^{i\varphi}|01\rangle\langle 0r|+\mathrm{H.c.})+\Delta|0r\rangle\langle 0r|,\\
H_{10}&=(\frac{\Omega}{2}e^{i\varphi}|10\rangle\langle r0|+\mathrm{H.c.})+\Delta|r0\rangle\langle r0|,\\\nonumber
H_{11}&=[\frac{\sqrt{2}}{2}\Omega e^{i\varphi}(|11\rangle\langle R|+|R\rangle\langle rr|)+\mathrm{H.c.}]\\
&+\Delta|R\rangle\langle R|+(V+2\Delta)|rr\rangle\langle rr|,
\end{align}
\end{subequations}
where $|R\rangle=(|1r\rangle+|r1\rangle)/\sqrt{2}$. The state $|00\rangle$ is decoupled from the other states.

We first consider the case of Rydberg blockade ($\Omega, \Delta\ll V$). Since the controlled-NOT gate can be transformed from the controlled-Z gate with two $\pi/2$-pulses applying to the target qubit \cite{Isenhower2010}, we only focus on the gate time of the controlled-Z gate. There are mainly two kinds of schemes to realize controlled-Z gate:

(i) Pulses sequences \cite{Saffman2016,Isenhower2010}. The $k$-th atom (control qubit) is first excited to $|r_k\rangle$ with $\pi$ pulse, and then a $2\pi$ pulse is applied to the $j$-th atom (target qubit), finally, a second $\pi$ pulse return the $k$-th atom to the ground state. One may find that $|00\rangle\rightarrow|00\rangle$, $|01\rangle\rightarrow-|01\rangle$, $|10\rangle\rightarrow-|10\rangle$, $|11\rangle\rightarrow-|11\rangle$ after the control. The total gate time $T_b$ is given by $\Omega T_b=4\pi$ which is irrelevant to the Rydberg interaction strength $V$.

(ii) Single modulated pulse \cite{Levine2019,Sun2020,Chen2024,Sun2023}. In such protocol a pulse with time-varying Rabi frequency/detuning/phase is apply to the two atoms simultaneously. The computational basis will evolve cyclically and gain a total phase $\varphi_b$ ($b=00, 01, 10, 11$) after the driving. The controlled phase is determined by $\varphi_c=\varphi_{11}-\varphi_{10}-\varphi_{01}$ which can be tilt through the optimized algorithm. The total gate time $T_b$ will be determined by $\Omega T_b=\eta\pi$, $\eta$ depends on the control scheme. Here $T_b$ is also irrelevant to the Rydberg interaction strength $V$.

Therefore, the gate time of two-qubit gate with Rydberg blockade is independent of the Rydberg interaction explicitly \cite{Note}. This can be understood by rewriting Eq.(5c) with Rydberg blockade condition which arrives at
\begin{equation}
H'_{11}=(\Omega e^{i\varphi}|R\rangle\langle b|+\mathrm{H.c.})+\Delta|R\rangle\langle R|,
\end{equation}
$|b\rangle=\sin(\theta_{11}/2)e^{-2i\varphi}|11\rangle+\cos(\theta_{11}/2)|rr\rangle$ with $\tan\theta_{11}=\Omega/\Delta$. As can be seen that $H'_{11}$ as well as the dynamics is independent of $V$. Physically, such results can be understood by realizing that the interactions between the atoms is by the aid of the Rydberg states, but not direct couplings. The increasing of Rydberg interaction strength will increase the interaction between the atoms, nevertheless, will also depress the probability of doubly excitation of Rydberg states (the doubly excitation of Rydberg states is the core to realize two-qubit Rydberg gate, though have been dropped, cannot be neglected).

\section{Two-qubit gates with weak Rydberg interactions}
In the general case, the gate time of two-qubit gate will be related to the interaction strength. Following the protocol in \cite{Ming2024}, geometric controlled-phase gate can be realized by using the pulse sequences as follow:
\begin{equation}
\begin{split}
&t: 0\rightarrow4T, \Omega=\kappa V, \Delta=-V/2;\\
&t: 0\rightarrow T, \varphi=0; t: T\rightarrow2T,  \varphi=\pi/2;\\
&t: 2T\rightarrow3T, \varphi=0; t: 3T\rightarrow4T, \varphi=\pi/2.
\end{split}
\end{equation}
One only needs to adjust the relative phases of Rydberg lasers at proper times and keeps the detuning and Rabi frequencies constant. The controlled phase $\varphi_c$ can be tilt by changing the ratio $\kappa$ between the Rabi frequencies and the interaction strength. Cyclic evolution is met when
\begin{equation}
\sqrt{4\Omega^2+V^2/4}T=2\pi,
\end{equation}
and thus the controlled-phase gate can be achieved. $\varphi_c=-\pi$ when $\kappa=1.65$, working at the region of weak Rydberg interaction.

In the following we make a comparison of the gate time between Rydberg blockade scheme and the weak interaction scheme under the same Rabi frequency $\Omega$. The gate time of Rydberg blockade scheme is given by $T_b=\eta\pi/\Omega$, $\eta>2$ due to the condition of cyclic evolution of the computational basis. The gate time of weak interaction scheme is given by $T_t=4T=8\pi/\sqrt{4\Omega^2+\Omega^2/(4\kappa^2)}$. $T_t=3.954\pi/\Omega$ when $\kappa=1.56$, which is near $T_b$. Actually, the gate time $T_t$ can be decreased by increasing $V$ if we are not limit to the controlled-Z gate. It can be found that $T_t<2\pi/\Omega$ when $\kappa=0.144$ ($V=6.93\Omega$) which is still far from the Rydberg blockade region. Therefore, the introduction of weak interaction condition does not mean a increase in the gate time.

Here we want to make a comparison among the geometric control (7) with other control schemes. It seems that pulses sequences (7) is similar to the Rydberg anti-blockade scheme since it requires $\Delta=-V/2$. However, the interaction strength $V$ of (7) is not needed to fulfill the blockade condition. The atomic spacing needs not to be small and the scheme will be insensitive to the thermal motion of the atoms. Discussion of gate fidelity versus the atomic thermal motions can be found in \cite{Ming2024}. It can be also found that the geometric control use the strategy of composite pulses which will enhance the robustness against the parameters variation, especially the Rabi frequency $\Omega$ \cite{Ming2024}.

Finally, by realizing two-qubit gate with weak Rydberg interaction, we address the following significant advantages as shown in \cite{Ming2024}:

(i) The scheme comes up with a new-type geometric quantum gate in the high dimensional Hilbert space. It consolidates the unconventional geometric phase and the non-adiabatic holonomic phase simultaneously. Furthermore, new-type geometric interferometers can be also investigated in such system.

(ii) The proposed two-qubit gates remove the restriction of Rydberg blockade and will reduce the crosstalk between the atoms significantly.

(iii) The cost of actuating quantity of the proposed gate is 90 times smaller than the Rydberg blockade scheme. Therefore, the abandon of Rydberg blockade does not mean a increase of gate time.

\section{Conclusion}
In summary, we have discussed the relationship between the gate time of Rydberg two-qubit gate and the Rydberg interaction strength. For the case of weak Rydberg interaction, the interaction will help to accelerate the two-qubit gate. This results that the scheme of weak Rydberg interaction will be faster than the Rydberg blockade scheme under the same Rabi frequencies of Rydberg excitation. The introduction of pulses sequences (7) and weak Rydberg interaction will significantly reduce the crosstalk between the atoms and the dephasing effect, furthermore, such scheme based on is robust against random noises due to the geometric characteristic. Such scheme provides a platform to investigate the interference between two different types of geometric phases and may trigger the investigation of new-type geometric gates in high-dimension Hilbert space (i.e., adiabatic geometric/adiabatic non-Abelian-mixed multi-qubit gates).


\begin{thebibliography}{99}
\bibitem{Note}Although increasing $V$ will induce a more relax blockade condition which allows a lareger $\Omega$ to shorten the gate time, is not what we want to discuss in this manuscript. Here we investigate the relationship between $T$ and $\Omega$ under a big enough and constant $V$.
\end{thebibliography}
\end{document}